\documentclass[11pt]{article}
\usepackage{amssymb,amsmath,amsfonts}
\usepackage{graphicx}
\usepackage{graphics}
\usepackage{eepic,epsfig}
\usepackage{cite}
\usepackage{hyperref}

\begin{document}
\title{Vacuum expectation value of the energy-momentum tensor in a higher dimensional  compactified cosmic string spacetime}
\author{E. A. F. Bragan\c{c}a$^1$\thanks{E-mail: eduardo.braganca@ufes.br/braganca@df.ufpe.br} ,
H. F. Santana Mota$^2$\thanks{E-mail: hmota@fisica.ufpb.br} \ and \ E. R. Bezerra de Mello$^3$\thanks
{E-mail: emello@fisica.ufpb.br}\\
\\
\textit{$^{1}$N\'{u}cleo Cosmo-Ufes e Departamento de F\'{i}sica, CCE,}\\
\textit{Universidade Federal do Esp\'{i}rito Santo, 29075-910,Vit\'{o}ria-ES, Brazil}\\
\textit{and}\\
\textit{Departamento de F\'{i}sica, Universidade Federal de Pernambuco,}\\
\textit{52171-900, Recife-PE, Brazil}
\vspace{%
0.3cm}\\
\textit{$^{2,3}$Departamento de F\'{i}sica, Universidade Federal da Para\'{i}ba,}\\
\textit{58.059-970, Caixa Postal 5.008, Jo\~{a}o Pessoa, PB, Brazil}}
\maketitle
%
\begin{abstract}
	The main objective of this paper is to analyze the vacuum expectation value (VEV) of the energy-momentum tensor (EMT) associated with a charged scalar quantum field in a high-dimensional cosmic string spacetime admitting the presence of a magnetic flux running along the string's core. In addition, we also assume that the coordinate along the string's axis is compactified to a circle and presents an extra magnetic flux running along its center. This compactification is implemented by imposing a quasiperiodic condition on the field with an arbitrary phase $\beta$. The calculation of the VEV of the EMT and field squared, are developed by using the positive-energy Wightman function. The latter is constructed  by the mode sum of the complete set of normalized bosonic wave-functions. Due to the compactification, two distinct contributions take place. The first one corresponds to the VEV in a cosmic string spacetime without compactification considering the magnetic interaction. So, this term presents a contribution due to the non-trivial topology of the conical space, and an additional contribution due to the interaction between the scalar field with the magnetic flux. The latter is a periodic function of the magnetic flux with period equal to the quantum flux, $\Phi_0=2\pi/e$, and corresponds to a Aharanov-Bhom type contribution. The second contribution is induced by the compactification itself and depends on the magnetic flux along the string's core. It is also an even function of the magnetic flux enclosed by the string axis. Some asymptotic expressions for the VEVs of the energy-momentum tensor and field squared are provided for specific  limiting cases of the physical parameter of the model.
\end{abstract}
\bigskip

%

\bigskip

\section{Introduction}
\label{Int}
Topological and geometrical concepts are of great importance in recent
developments of many areas of physics, including condensed matter physics,
gauge field theories and cosmology. In particular, the global properties of the
spacetime play an important role in quantum field theory. Many interesting quantum effects arise as consequence of the nontrivial topological structure of
spacetime. A well-known example of such quantum phenomena is the
topological Casimir effect (for reviews see \cite{Eliz94,Most,Milton,Bord,Bord02}). This
effect is one of the most significant macroscopic manifestations of quantum
properties of the vacuum state. The periodicity conditions imposed on a
quantum field due to nontrivial topology lead to a modification of the zero-point fluctuations resulting in the shifting in the vacuum expectation values (VEVs) for physical quantities. In Kaluza-Klein type models, the dependence of the vacuum energy on the lengths of extra
dimensions can serve as a mechanism for the stabilization of moduli fields.
 The topological Casimir effect in cylindrical and toroidal carbon nanotubes is investigated in \cite{Bell09,Eliz11} within the framework of a Dirac-like theory for the electronic states in graphene. In addition to fermionic field, the scalar and gauge fields originated from the elastic properties and describing disorder phenomena, like the distortion of graphene lattice and structure defects should be taking into consideration \cite{Jackiw,Oliveira}. In graphene made structures, like cylindrical and toroidal carbon
nanotubes, the background geometry for the corresponding field theory
contains one or two compact dimensions. In quantum field theory, the
periodicity conditions imposed on the field operator along compact
dimensions modify the spectrum for the normal modes and as a result
the vacuum expectation values of physical observables are changed.

Cosmic strings are lines of trapped energy density, analogous to defects such as vortex lines in superconductors and superfluids.  An enormous number of string type solutions have been found in many different field theoretic models, including electroweak strings in Weinberg-Salam
model and strings in grand unified theories \cite{CopAndPogo,CandK,Pol,Jeannerot}. 
They can curve spacetime and be of cosmological
and astronomical significance, contributing to a large number of phenomena. For instance, cosmic strings can generate anisotropies
in the cosmic microwave background, non-Gaussianity and B-mode polarization. They can also be source of gravitational waves and produce high energy cosmic rays and gravitationally lensing astrophysical objects
\cite{Hindmarsh2,KandMandNaka,Virb,FraAndRing,PogoAndWym,VirbAndEllis,LaixAndKrauss,KaiserAndSte,AVile}. 

The geometry of the spacetime associated with an idealized cosmic string,  i.e., infinitely long and straight, is locally flat but topologically conical, having a planar angle deficit given by $\Delta\phi=8\pi G\mu_0$
on the two-surface orthogonal to the string, where $G$
is Newton's constant and $\mu_0$ is the mass per unit length, proportional to the square of the symmetry breaking scale.

Although the geometry of the spacetime produced by an idealized cosmic
string is locally flat, its conical structure modifies the vacuum fluctuations associated with quantum fields. As a consequence, a non-vanishing vacuum expectation value (VEV)  of physical observables like the energy-momentum tensor (EMT), $\langle T_{\mu\nu}\rangle$, takes place. The calculation of the VEV of the EMT associated with the scalar and fermionic fields in the idealized cosmic string spacetime has been developed in  \cite{PhysRevD.35.536, escidoc:153364, GL, DS, PhysRevD.46.1616}
and \cite{PhysRevD.35.3779, LB, Moreira1995365, BK}, respectively. Furthermore, the presence of a magnetic flux running through the core of the string gives additional contributions to the VEVs associated with charged fields
\cite{PhysRevD.36.3742, guim1994, SBM, SBM2, SBM3, Spinelly200477, SBM4}. 

 In the present paper we would like to extend the analysis of the VEV of the EMT associated with charged scalar quantum field in the cosmic string spacetime. In order to do that we include two important ingredients: We assume arbitrary value for the planar angle deficit, and compactified the dimension  along the string's axis in a circle enclosing an extra magnetic flux. In this way we intend to complete the analysis of a scalar quantum field in this background, started in Refs. \cite{Mello12,Braganca,Braganca02,azadeh}.
 
The plan of this paper is the following: in section \ref{Wightman} we present the geometry background spacetime that we want to work. Also we provide the complete set of normalized wave-function associated with the scalar quantum field obeying the quasiperiodicity along the direction of the string's core and in the presence of two different magnetic fluxes. By using the mode sum formula,  we calculate the positive-energy Wightman function. We explicitly show that this function is composed by two parts: The first one is due to the pure cosmic string spacetime without compactification, and the second is induced by the compactification.  In section \ref{Sec.phi2}, we evaluated the renormalized VEV of the field squared. This expectation value that is formally given by taking the coincidence limit of the Wightman function, is also decomposed into two parts. The VEV of the energy-momentum tensor is presented in section \ref{Sec.EMT}, where it is explicitly show to obey the conservation condition. Finally, the most relevant results are summarized in section \ref{Concl}.

\section{Wightman function}
\label{Wightman}
%

In this section we present first the geometry of the spacetime that we want to work. It corresponds to a  generalization of a four-dimensional idealized cosmic string spacetime. Considering $D$ as the dimension of the space section, with $D\geq 3$. The geometry of this $(D+1)-$dimensional conical space is given by the line element below,
\begin{equation}
ds^{2}=g_{\mu\nu}dx^{\mu}dx^{\nu}=dt^{2}-dr^2-r^2d\phi^2-dz^2- \sum_{i=4}^{D}(dx^{i})^2 \ ,
\label{eq01}
\end{equation}
where we are making use of the generalized cylindrical coordinates.

In this coordinate system we are assuming: $r\geq 0$, $0\leq\phi\leq 2\pi/q$ and $-\infty< (t, \ x^i) < +\infty$ for $i=4,...,D$.  
The presence of the cosmic string is seen through the presence of the parameter
$q\geq 1$. Moreover, we assume that the direction
along the $z$-axis is compactified to a circle with the length $L$, so $0\leqslant z\leqslant L$.
The standard cosmic string space-time is characterized by $D=3$, with $z\in(-\infty, \ \infty)$.
In this case $q^{-1}=1-4\mu_0$, being $\mu_0$ the linear mass density of the string.

The quantum dynamics of a charged bosonic field with mass $m$, in a curved background and in the presence of an electromagnetic potential vector, $A_\mu$, is governed by the equation
\begin{equation}
\left[\frac{1}{\sqrt{|g|}}D_\mu\left(\sqrt{|g|}\,g^{\mu\nu}D_\nu\right)+m^{2}+\xi R\right] \varphi (x)=0 \ ,
\label{eq02}
\end{equation}
where $D_{\mu}=\partial_{\mu}+ieA_{\mu}$ and $g={\rm det}(g_{\mu\nu})$. 
Considering a thin and infinitely straight cosmic string, we have that $R=0$ for $r\neq 0$. 

The main objective of this section is to present the solution of \eqref{eq02}, by imposing the quasiperiodicity condition on the matter field
\begin{equation}
\varphi (t,r,\phi,z+L,x^4,...,x^D)=e^{2\pi i\beta}\varphi(t,r,\phi,z,x^4,...,x^D) \ ,  
\label{eq03}
\end{equation}
with a constant phase $\beta $, $0\leqslant \beta \leqslant 1$, and considering the presence of a two independent vector potentials,
\begin{equation}
A_{\mu}=(0,0,{-q\Phi_\phi}/{2\pi},{-\Phi_z}/{L})\ , 
\label{eq04}
\end{equation}
being $\Phi_\phi$ and $\Phi_z$
the magnetic fluxes along the string's core and enclosed by the compactified direction, respectively.

The general solution of the Eq. \eqref{eq02} takes the form
\begin{equation}
\varphi_{\sigma}(x)=CJ_{q|n+\alpha|}(\lambda r)
e^{-i\omega t+iqn\phi+ik_z z+i{\vec{k}\cdot{\vec{r}}_{\parallel}}} \ , 
\label{eq05}
\end{equation}
with $C$ being a normalization constant and
\begin{eqnarray}
\lambda&=&\sqrt{\omega^2-{\vec{k}}^{2}-{\tilde{k}}_z^2-m^2} \ , \nonumber\\  
{\tilde{k}}_z&=&k_z+eA_z \ ,\nonumber\\
\alpha&=&\frac{eA_\phi}{q}=-\frac{\Phi_\phi}{\Phi_0} \ ,
\label{eq06}
\end{eqnarray}
with $\Phi_0=2\pi/e$ being the quantum flux.
The quantum number $k_z$ is discretized by the Eq. \eqref{eq03} as
\begin{equation}
k_z=k_l=\frac{2\pi}{L}(l+\beta) \ \ {\rm with} \  l=0\ ,\pm 1,\pm 2,... \ .
\label{eq07}
\end{equation}

Take this into account, the energy assumes the form
\begin{equation}
\omega=\omega_l=\sqrt{m^2+\lambda^2+{\tilde{k}}^2_l+{\vec{k}}^2} \ , 
\label{eq08}
\end{equation}
where
\begin{eqnarray}
{\tilde{k}}_z&=&{\tilde{k}}_l=\frac{2\pi}{L}(l+\tilde{\beta}) \ , \nonumber\\
\tilde{\beta}&=&\beta+\frac{eA_zL}{2\pi}=\beta-\frac{\Phi_z}{\Phi_0} \ .
\label{eq09}
\end{eqnarray}

The normalized bosonic wave-function reads \cite{Braganca},
\begin{equation}
\varphi_{\sigma}(x)=\left[\frac{q\lambda}{2(2\pi)^{D-2}\omega_l L}\right]^{\frac{1}{2}} 
J_{q|n+\alpha|}(\lambda r)e^{-i\omega t+iqn\phi+ik_l z+i{\vec{k}\cdot{\vec{r}}_{\parallel}}} \ .
\label{eq10}
\end{equation}

To analyze the field squared and the energy-momentum tensor we need to find the positive
frequency Wightman function
$W(x,x')=\left\langle 0|\varphi(x)\varphi^{*}(x')|0 \right\rangle$, 
where $|0 \rangle$ stands for the vacuum state. In this context the Wightman function is presented in terms of the mode sum formula below,
\begin{equation}
W(x,x')=\sum_{\sigma}\varphi_{\sigma}(x)\varphi_{\sigma}^{*}(x') \ ,
\label{eq11}
\end{equation}
where we are using the compact notation defined as
\begin{equation}
\sum_{\sigma }=\sum_{n=-\infty}^{+\infty} \ \int d{\vec{k}} \ \int_0^\infty
\ d\lambda \ \sum_{l=-\infty }^{+\infty} \ .  \label{Sumsig}
\end{equation}

Substituting \eqref{eq10} into the sum \eqref{eq11} we obtain
\begin{eqnarray}
W(x,x')&=&\frac{q}{2L(2\pi)^{D-2}}\sum_\sigma e^{iqn\Delta\phi} 
 e^{i{\vec{k}\cdot{\Delta\vec{r}}_{\parallel}}} \ 
\lambda J_{q|n+\alpha|}(\lambda r) J_{q|n+\alpha|}(\lambda r')\nonumber\\
&\times&\frac{e^{-i\omega_l\Delta t+ik_l\Delta z}}{\omega_l} \ ,
\label{eq12}
\end{eqnarray}
where $\Delta\phi=\phi-\phi'$, $\Delta{\vec{r}}_{\parallel}={\vec{r}}_{\parallel}-
{\vec{r}}_{\parallel}'$, $\Delta t=t-t'$ and $\Delta z=z-z'$. 

The  mode functions in Eq. \eqref{eq12} are specified by the set of quantum numbers 
$\sigma=\{n,\lambda,k_l, {\vec{k}}\}$, with the values in the ranges $n=0,\pm1,\pm2, \ \cdots$, 
$-\infty<k^j<+\infty$ with $j=4, \cdots \ D$, $0<\lambda<\infty$ and $k_l=2\pi(l+\beta)/L$
with $l=0,\pm1,\pm2, \ \cdots$. 

For the evaluation of the sum over the quantum number \textit{l} we use the Abel-Plana Summation formula given in the form \cite{PhysRevD.82.065011}
\begin{eqnarray}
\sum_{l=-\infty }^{\infty }g(l+\tilde{\beta} )f(|l+\tilde{\beta} |)&=&\int_{0}^{\infty }du
\left[ g(u)+g(-u)\right] f(u) + i\int_{0}^{\infty }du\left[ f(iu)-f(-iu)\right]\nonumber\\
&\times&\sum_{j =\pm1}\frac{g(ij u)}{e^{2\pi (u+ij \tilde{\beta} )}-1}.
\label{sumform}
\end{eqnarray}

In Eq. \eqref{eq12} we can write the exponential dependence of $k_l$ as
\begin{equation}
e^{ik_z \Delta z}=e^{2\pi i(l+\beta)\Delta z/L}=e^{i\tilde{k}_z \Delta z}e^{2\pi i\Phi_z \Delta z/(L\Phi_0)}=
e^{i\tilde{k}_z \Delta z}\delta(\Delta z),
\end{equation}
with $\delta(\Delta z)=e^{- ie A_z \Delta z}$.
Using the above equation in \eqref{eq12} allows us to use the Abel-Plana summation formula taking
\begin{equation}
g(u)= e^{(2\pi i/{L}) \Delta z} \ \ \ {\rm and} \ \ \ f(u)=\frac{e^{-i\Delta t\sqrt{m^2+\lambda^2 +k^2+(\frac{2\pi u}{L})^2}}}
{\sqrt{m^2+\lambda^2 +k^2+(\frac{2\pi u}{L})^2}},
\end{equation}
with $u=l+\tilde\beta$. Doing this, we can write the Wightman function in the decomposed form
\begin{equation}
W(x,x')=W_{cs}(x,x')+W_c(x,x'),
\label{decomposed}
\end{equation}
where $W_{cs}(x,x')$ is the contribution in the cosmic string background without the compactification coming from the first term
on the right hand side of Eq. \eqref{sumform}, and $W_{c}(x,x')$ is the contribution due the compactification and comes from the second term on the right hand side.

For the contribution due the cosmic string background, we have
\begin{eqnarray}
W_{cs}(x,x')&=&\frac{q\delta(\Delta z)}{(2\pi)^{D-2}L}\sum_{n=-\infty}^\infty e^{inq\Delta\phi}\int d\vec{k} \ e^{i\vec{k}\cdot \Delta\vec{r}_{||}}
\int_0^\infty d\lambda \ \lambda J_{q|n+\alpha|}(\lambda r)J_{q|n+\alpha|}(\lambda r')\nonumber\\
&\times&\int_0^\infty du \cos(2\pi u\Delta z/L)\frac{e^{-i\Delta t\sqrt{(2\pi u/L)^2 + a^2}}}{\sqrt{(2\pi u/L)^2 + a^2}},
\label{eq13}
\end{eqnarray}
where $a^2=m^2+\lambda^2 +\vec{k}^2$.
Introducing a new variable $y=2\pi u/L$ and taking the Wick rotation, $\Delta t = i\Delta\tau$, we are able to use the following identity
\begin{equation}
\frac{e^{\omega_l \Delta\tau}}{\omega_l}=\frac{2}{\pi}\int_0^\infty ds \ e^{-\omega_l^2s^2-\Delta\tau^2/4s^2},
\label{eq14}
\end{equation}
and solve the integral over $y$ and $\vec{k}$. After these steps one finds
\begin{eqnarray}
W_{cs}(x,x')&=&\frac{q\delta(\Delta z)}{2^{D-1}\pi^{(D+1)/2}}\sum_{n=-\infty}^{\infty}e^{inq\Delta \phi}
\int_0^\infty ds \ \frac{e^{-m^2s^2-((\Delta \tau)^2+(\Delta z)^2+|\Delta\vec{r}_{||}|^2)/4s^2}}{s^{D-2}}\nonumber\\
&\times&\int_0^\infty d\lambda \ \lambda J_{q|n+\alpha|}(\lambda r)J_{q|n+\alpha|}(\lambda r')e^{-\lambda^2s^2}.
\label{eq15}
\end{eqnarray}
To perform the integration over the quantum number $\lambda$ we use the formula \cite{prudnikov2}
\begin{equation}
\int_0^\infty d\lambda \ \lambda J_{\beta_n}(\lambda r)J_{\beta_n}(\lambda r')e^{-s\lambda^2}=\frac{1}{2s}
\mathrm{exp}\left(-\frac{r^2+r'^2}{4s}\right)I_{\beta_n}\left(\frac{rr'}{2s}\right).
\label{int_lambda}
\end{equation}
Defining a new variable $u=1/2s^2$, we find
\begin{eqnarray}
W_{cs}(x,x')&=&\frac{q\delta(\Delta z)}{2(2\pi)^{(D+1)/2}}\int_0^\infty du \ u^{\frac{D-3}{2}}e^{-\frac{m^2}{2u}-\frac{u}{2}(r^2+r'^2+|\Delta\vec{r}_{||}|^2
+(\Delta z)^2- (\Delta t)^2)} \ \mathcal{I}(\alpha,\Delta\phi,urr'),\nonumber\\
\label{eq16}
\end{eqnarray}
where we have introduced
\begin{equation}
\mathcal{I}(\alpha,\Delta\phi,urr')=\sum_{n=-\infty}^\infty e^{inq\Delta\phi}I_{q|n+\alpha|}(urr') \  .
\label{nsum}
\end{equation}

We can obtain a more workable expression for \eqref{eq16} writing the parameter $\alpha$ in \eqref{eq06} in the form
\begin{equation}
\alpha=n_0 +\alpha_0 \ \ {\rm with  \ |\alpha_0| \ < 1/2},
\label{alphazero}
\end{equation}
where $n_0$ is an integer number. In addition we use the form of the summation over $n$ given in \cite{deMello:2014ksa}, that we reproduce below:
\begin{eqnarray}
\mathcal{I}(\alpha,\Delta\phi,x)&=&\frac{1}{q}\sum_k e^{x\cos(2k\pi /q-\Delta\phi)}e^{i\alpha(2k\pi -q\Delta\phi)}-
\frac{e^{-iqn_0\Delta\phi}}{2\pi i}\sum_{j=\pm 1}je^{ji\pi q\alpha_0}\nonumber\\
&\times&\int_0^\infty dy \frac{\cosh[qy(1-\alpha_0)]-\cosh(q\alpha_0y)e^{-iq(\Delta\phi+j\pi)})}{e^{x\cosh y}[\cosh(qy)-\cos(q(\Delta\phi+j\pi))]}.
\label{representation}
\end{eqnarray}
For the summation over $k$ we have the condition
\begin{equation}
-\frac{q}{2}+\frac{2\pi}{q}\Delta\phi \leq k \leq \frac{q}{2}+\frac{2\pi}{q}\Delta\phi.
\label{conditionk}
\end{equation}

Using \eqref{representation} with $x=urr'$, the contribution for the Wightman function is the background of a cosmic string without compactification
is given by
\begin{eqnarray}
W_{cs}(x,x')&=&\frac{m^{D-1}\delta(\Delta z)}{(2\pi)^{\frac{D+1}{2}}}\left\{\sum_k e^{i\alpha(2k\pi-q\Delta\phi)}
f_{\frac{D-1}{2}}\left[m\sqrt{u_0^2-2rr'\cos(2k\pi/q-\Delta\phi)}\right]\right.\nonumber\\
&-&\left.\frac{qe^{-iqn_0\Delta\phi}}{2\pi i}\sum_{j=\pm1}je^{ij\pi q\alpha_0}\int_0^\infty dy
\frac{\cosh[qy(1-\alpha_0)]-\cosh(q\alpha_0y)e^{-iq(\Delta\phi+j\pi)}}{\cosh(qy)-\cos[q(\Delta\phi+j\pi)]}\right.\nonumber\\
&\times&\left. f_{\frac{D-1}{2}}\left[m\sqrt{u_0^2+2rr'\cosh y}\right]
\right\},
\label{wcs}
\end{eqnarray}
with $u_0^2=r^2+r'^2+|\Delta\vec{r}_{||}|^2+(\Delta z)^2- (\Delta t)^2$. In the above equation we have defined
\begin{equation}
f_\nu (x)=\frac{K_\nu (x)}{x^\nu}.
\label{eq17}
\end{equation}

The second term in the right hand side of the Abel-Plana summation formula, introducing $y=2\pi u/L$, give us
\begin{eqnarray}
W_c(x,x')&=&\frac{q\delta(\Delta z)}{(2\pi)^{D-1}} 
\sum_{n=-\infty}^{+\infty}e^{inq\Delta \phi}\int d\mathbf{k} \ e^{i{\mathbf{k}}\cdot \Delta \mathbf{r}_{||}}
\int_{0}^{\infty}d\lambda \ \lambda J_{q|n+\alpha|}(\lambda r)J_{q|n+\alpha|}(\lambda r')\nonumber \\
&\times&\int_{\sqrt{m^2+\lambda^2+k^2}}^{\infty}dy\frac{\cosh({\Delta t\sqrt{y^2-m^2-\lambda^2-k^2}})}{\sqrt{y^2-m^2-\lambda^2-k^2}}
\sum_{j=\pm 1}\frac{e^{-yj\Delta z}}{e^{Ly+2\pi ij\tilde{\beta}}-1}.
\label{eq18}
\end{eqnarray}

We can solve the integration over the angular part of  $\mathbf{k}$ with the help of the formula
\begin{equation}
\int d\mathbf{k} \ e^{i\mathbf{k}|\Delta\mathbf{r}_{||}|}F(k)=\frac{(2\pi)^{\frac{D-3}{2}}}{|\Delta \mathbf{r}_{||}|^{\frac{D-5}{2}}}
\int_0^\infty dk \ k^{(D-3)/2}J_{\frac{D-5}{2}}(k|\Delta\mathbf{r}_{||}|)F(k),
\end{equation}
for a given function $F(k)$. 
Using the expansion $(e^u-1)^{-1}=\sum_{l=1}^\infty e^{-lu}$ and solving the integral over $y$ we find
\begin{eqnarray}
W_c(x,x')&=&\frac{q\delta(\Delta z)}{(2\pi)^{\frac{D+1}{2}}}\sideset{}{'}\sum_{l=-\infty}^{\infty}e^{-2\pi il\tilde{\beta}}
\int_0^\infty dk\frac{k^{(D-3)/2}}{|\Delta\mathbf{r}_{||}|^{(D-5)/2}}J_{(D-5)/2}(k|\Delta\mathbf{r}_{||}|)\sum_{n=-\infty}^\infty e^{inq\Delta\phi}\nonumber\\
&\times&\int_0^\infty d\lambda \ \lambda J_{q|n+\alpha|}(\lambda r)J_{q|n+\alpha|}(\lambda r')
K_0\left(\sqrt{(\lambda^2+m^2+k^2)(\Delta z+lL)^2-(\Delta t)^2}\right).\nonumber\\
\label{eq19}
\end{eqnarray}

The prime in the summation over $l$ means that the term $l=0$ must be excluded.
For further manipulations of the above expression, we employ the integral representation of the
modified Bessel function \cite{watson1995treatise}
\begin{equation}
K_\nu(x)=\frac{1}{2^{\nu+1}x^\nu}\int_0^\infty d\tau\frac{e^{-\tau x^2-1/(4\tau)}}{\tau^{\nu+1}}  \  , 
\label{Kfunction}
\end{equation}
which allow us to perform the integration over
$k$ using the formula from \cite{prudnikov2} and the integration over $\lambda$ using \eqref{int_lambda}.
Introducing a new variable $u=1/(2\tau b^2)$, with $b^2=(\Delta z+lL)^2-(\Delta t)^2$, and changing $l\rightarrow -l$ one finds
\begin{eqnarray}
W_c(x,x')&=&\frac{q\delta(\Delta z)}{2(2\pi)^{(D+1)/2}}\sideset{}{'}\sum_{l=-\infty}^{\infty}e^{2\pi il\tilde{\beta}}
\int_0^\infty du \ u^{\frac{D-3}{2}}e^{-\frac{m^2}{2u}-\frac{uu^2_l}{2}}\ \mathcal{I}(\alpha,\Delta\phi,urr'),
\label{eq20}
\end{eqnarray}
with the function introduced by Eq. \eqref{nsum} and
\begin{equation}
u^2_l=r^2+r'^{2}+(\Delta z -lL)^2+|\Delta \vec{r}_{||}|^2-(\Delta t)^2,
\label{ul}
\end{equation}

By using Eq. \eqref{representation}, with $x=urr'$, in \eqref{eq20} and taking into account \eqref{alphazero}, the integration
over the variable $u$ can be directly solved, and we finally obtain
\begin{eqnarray}
W_c(x,x')&=&\frac{m^{D-1}\delta(\Delta z)}{(2\pi)^{\frac{D+1}{2}}}\sideset{}{'}\sum_{l=-\infty}^\infty e^{2\pi il\tilde{\beta}}\left\{\sum_k e^{i\alpha(2k\pi-q\Delta\phi)}
f_{\frac{D-1}{2}}\left[m\sqrt{u_l^2-2rr'\cos(2k\pi/q-\Delta\phi)}\right]\right.\nonumber\\
&-&\left.\frac{qe^{-iqn_0\Delta\phi}}{2\pi i}\sum_{j=\pm1}je^{ij\pi q\alpha_0}\int_0^\infty dy
\frac{\cosh[qy(1-\alpha_0)]-\cosh(q\alpha_0y)e^{-iq(\Delta\phi+j\pi)}}{\cosh(qy)-\cos[q(\Delta\phi+j\pi)]}\right.\nonumber\\
&\times&\left. f_{\frac{D-1}{2}}\left[m\sqrt{u_l^2+2rr'\cosh y}\right]
\right\}.
\label{eq21}
\end{eqnarray}

We can note that the missing $l=0$ term in the above equation is the one given by Eq. \eqref{wcs}. So, the total Wightman function is written as
\begin{eqnarray}
W(x,x')&=&\frac{m^{D-1}\delta(\Delta z)}{(2\pi)^{\frac{D+1}{2}}}\sum_{l=-\infty}^\infty e^{2\pi il\tilde{\beta}}\left\{\sum_k e^{i\alpha(2k\pi-q\Delta\phi)}
f_{\frac{D-1}{2}}\left[m\sqrt{u_l^2-2rr'\cos(2k\pi/q-\Delta\phi)}\right]\right.\nonumber\\
&-&\left.\frac{qe^{-iqn_0\Delta\phi}}{2\pi i}\sum_{j=\pm1}je^{ij\pi q\alpha_0}\int_0^\infty dy
\frac{\cosh[qy(1-\alpha_0)]-\cosh(q\alpha_0y)e^{-iq(\Delta\phi+j\pi)}}{\cosh(qy)-\cos[q(\Delta\phi+j\pi)]}\right.\nonumber\\
&\times&\left. f_{\frac{D-1}{2}}\left[m\sqrt{u_l^2+2rr'\cosh y}\right]
\right\}.
\label{wightmanfinal}
\end{eqnarray}
This is the final and more compact expression for the total Wightman function and it allows us to present the VEVs of the field squared and EMT for a massive scalar field in a closed form for a general value of $q$.


\section{Vacuum expectation value (VEV) of the field squared}
\label{Sec.phi2}

We can obtain the VEV of the field squared taking the coincidence limit of the arguments in Eq. \eqref{wightmanfinal}. This provides its expression in the
decomposed form
\begin{equation}
\langle|\varphi|^2\rangle=\langle|\varphi|^2\rangle_{cs}+\langle|\varphi|^2\rangle_c.
\end{equation}
where $\langle|\varphi|^2\rangle_{cs}$ corresponds to the geometry of a cosmic string without compactification
and $\langle\varphi^2\rangle_c$ correspond to the contribution induced by the compactification along
the $z$-direction. When we take the coincidence limit, we note that the first contribution in the right hand side of the above
equation is divergent and only the term $\langle|\varphi|^2\rangle_c$ is finite. To obtain a finite and well defined result to $\langle|\varphi|^2\rangle_{cs}$, we have to renormalize it by subtracting the corresponding Wightman function in the Minkowiski spacetime, as show below:
\begin{equation}
\label{phiren}
\langle|\varphi|^2\rangle_{cs}= \lim_{x'\rightarrow x} [W_{cs}(x,x')- W_M(x,x')] \  .
\end{equation}
This procedure can be done in a manifest form by discarding the $k=0$ component of the Wightman function, Eq. \eqref{wcs}. As a result we find
\begin{eqnarray}
\label{phi2_cs}
\langle|\varphi|^2\rangle_{cs}&=&\frac{2m^{D-1}}{(2\pi)^{\frac{D+1}{2}}}\left\{\sum_{k=1}^{[q/2]}\cos(2k\pi \alpha_0)f_{\frac{D-1}{2}}(2mr s_k)
-\frac{q}{\pi}\int_0^\infty dy \frac{g(q,\alpha_0,2y)}{\cosh(2qy)-\cos(q\pi)}\right.\nonumber\\
&\times&\left. f_{\frac{D-1}{2}}(2mr c_y)\right\} \  ,
\end{eqnarray}
where $[q/2]$ stands for the integer parte of $q/2$, and
\begin{equation}
g(q,\alpha_0,2y)=\sin(q\pi|\alpha_0|)\cosh[2qy(|\alpha_0|-1)]-\cosh(2q\alpha_0 y)\sin[q\pi(|\alpha_0|-1)]  \ ,
\label{gfunction}
\end{equation}
with
\begin{eqnarray}
s_k=\sin(k\pi/q) \ {\rm and} \ c_y=\cosh(y) \  .
\end{eqnarray}
For $1\leq q < 2$, the first term in the square brackets is absent.

For the massless field, the cosmic string contribution for the field squared is given by
\begin{equation}
\label{phi2massless}
\langle|\varphi|^2\rangle_{cs}=\frac{\Gamma(\frac{D-1}{2})}{2^{D}\pi^{\frac{D+1}{2}}r^{D-1}}
g_D(q,\alpha_0),
\end{equation}
where the function $g_D(q,\alpha_0)$ is defined as
\begin{equation}
g_D(q,\alpha_0)=\sum_{k=1}^{[q/2]}\frac{\cos(2\pi k\alpha_0)}{s_k^{D-1}}
-\frac{q}{\pi }\int_{0}^{\infty }dy\frac{g(q,\alpha_0,2y)}{\cosh (2qy)-\cos (q\pi)}c_y^{1-D}.
\label{gD}
\end{equation}

Considering the field squared in the limit which $mr\gg 1$, we obtain
\begin{eqnarray}
\label{phi2mr}
\langle|\varphi|^2\rangle_{cs}&=&\frac{m^{D/2-1}}{(4\pi r)^{D/2}}\left[\sum_{k=1}^{[q/2]}\frac{\cos(2k\pi\alpha_0)}{s_k^{D/2}}e^{-2mrs_k}
-\frac{q}{\pi}\int_0^\infty dy \frac{g(q,\alpha_0,2y) c_y^{-D/2}}{\cosh(2qy)-\cos(q\pi)}e^{-2mrc_y}\right].\nonumber\\
\end{eqnarray}

For the contribution induced by the compactification along the $z-$axis we directly obtain from \eqref{wightmanfinal}
\begin{eqnarray}
\label{phi2_c}
\langle|\varphi|^2\rangle_{c}&=&\frac{4m^{D-1}}{(2\pi)^{\frac{D+1}{2}}}\sum_{l=1}^\infty \cos(2\pi l\tilde{\beta})
\left\{\sideset{}{'}\sum_{k=0}^{[q/2]}\cos(2k\pi \alpha_0) f_{\frac{D-1}{2}}\left(m\sqrt{4r^2s_k^2+(lL)^2}\right)\right.\nonumber\\
&-&\left. \frac{q}{\pi}\int_0^\infty dy \frac{g(\alpha_0,q,2y)}{\cosh(2qy)-\cos(q\pi)}
f_{\frac{D-1}{2}}\left(m\sqrt{4r^2c_y^2 +(lL)^2}\right)\right\} \ ,
\end{eqnarray}
where the prime on the sign of the summation means that the $k = 0$ term should be divided by two. We can note that the above expression is finite on the string, i.e, $r=0$.

In the case of a massless field, the above equation becomes
\begin{eqnarray}
\langle|\varphi|^2\rangle_{c}&=&\frac{\Gamma(\frac{D-1}{2})}{\pi^{\frac{D+1}{2}}}\sum_{l=1}^\infty \cos(2\pi l\tilde{\beta})
\left\{ \sideset{}{'}\sum_{k=0}^{[q/2]}\frac{\cos(2k\pi \alpha_0)} {[4r^2s^2_k+(lL)^2]^{(D-1)/2}}\right.\nonumber\\
&-&\left. \int_0^\infty dy\frac{g(\alpha_0,q,2y)}{\cosh(2qy)-\cos(q\pi)}
[4r^2c_y^2+(lL)^2]^{(1-D)/2}\right\}.
\end{eqnarray}

We also can write the Eq. \eqref{phi2_c} as
\begin{equation}
\langle|\varphi|^2\rangle_{c}=\langle|\varphi|^2\rangle_{c}^{(0)}+\langle|\varphi|^2\rangle_{c}^{(q,\alpha_0)}.
\label{phi2_decomp}
\end{equation}
We note that the first term in the right-hand side of the above equation does not depend on $\alpha_0$ and $q$, being a
pure topological term. This term corresponds to the $k=0$ term of the summation over $k$ and is consequence only
of the compactification along the $z$-axis being given by
\begin{equation}
\langle|\varphi|^2\rangle_{c}^{(0)}=\frac{2m^{D-1}}{(2\pi)^{\frac{D+1}{2}}}\sum_{l=1}^\infty \cos(2\pi l\tilde{\beta})f_{\frac{D-1}{2}}(lmL) \ .
\label{phi0}
\end{equation}
Considering the limit $mL\ll1$, this pure topological term is written as
\begin{equation}
\langle|\varphi|^2\rangle_{c}^{(0)} \approx \frac{\Gamma(\frac{D-1}{2})}{2\pi^{\frac{D+1}{2}}L^{D-1}}\sum_{l=1}^{\infty}
\frac{\cos(2\pi l\tilde{\beta})}{l^{D-1}}.
\end{equation}
And in the limit $mL\gg1$ the main contribution comes from the $l=1$ term, which is given by
\begin{equation}
\langle|\varphi|^2\rangle_{c}^{(0)} \approx \frac{m^\frac{D-2}{2}}{(2\pi L)^{D/2}}e^{-mL}\cos(2\pi\tilde{\beta}) \  .
\label{Phi_1}
\end{equation}

The second term in the right hand side of Eq. \eqref{phi2_decomp} presents a dependence in the magnetic flux and planar
angle deficit, and is given by
\begin{eqnarray}
\label{phi2_decomp2}
\langle|\varphi|^2\rangle_{c}^{(q,\alpha_0)}&=&\frac{4m^{D-1}}{(2\pi)^{\frac{D+1}{2}}}\sum_{l=1}^\infty \cos(2\pi l\tilde{\beta})
\left\{\sum_{k=1}^{[q/2]}\cos(2k\pi \alpha_0) f_{\frac{D-1}{2}}\left(m\sqrt{4r^2s_k^2+(lL)^2}\right)\right.\nonumber\\
&-&\left. \frac{q}{\pi}\int_0^\infty dy \frac{g(\alpha_0,q,2y)}{\cosh(2qy)-\cos(q\pi)}
f_{\frac{D-1}{2}}\left(m\sqrt{4r^2c_y^2 +(lL)^2}\right)\right\}.
\end{eqnarray}
In the limit of $mL\gg1$ the main contribuition for the above equation comes from the $l=1$ term written as
\begin{eqnarray}
\langle|\varphi|^2\rangle_{c}^{(q,\alpha_0)} \approx \frac{2m^{\frac{D-2}{2}}}{(\pi L)^{D/2}}
\cos(2\pi\tilde{\beta})e^{-mL}\left[\sum_{k=1}^{[q/2]}\cos(2k\pi\alpha_0)-\frac{q}{\pi}
\int_0^\infty dy \frac{g(\alpha_0,q,2y)}{\cosh(2qy)-\cos(q\pi)}\right] \  .
\label{Phi_2}
\end{eqnarray}

\begin{figure}[h]
	\centering
	{\includegraphics[width=0.495\textwidth]{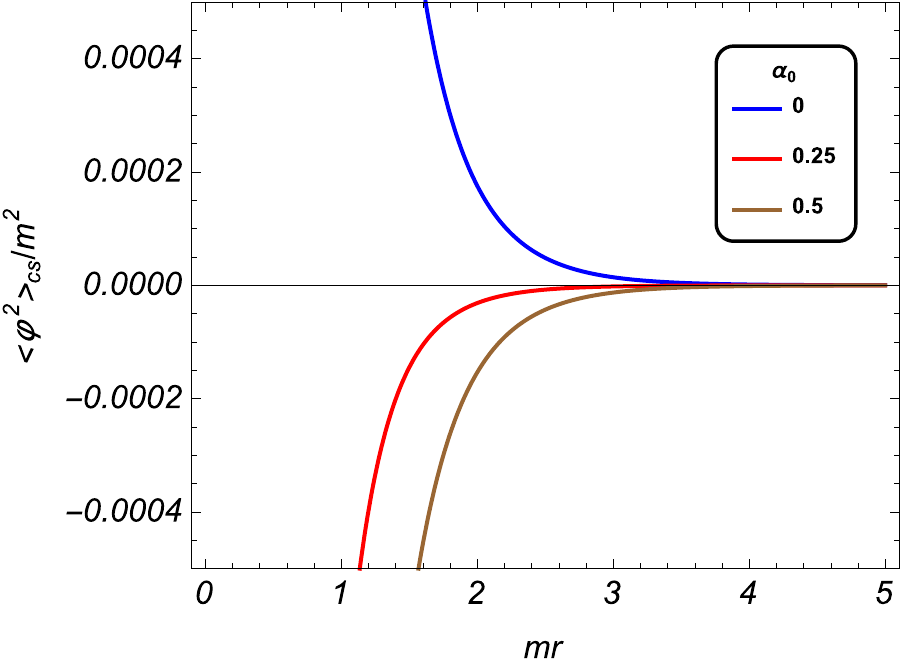}}
	\hfill
	{\includegraphics[width=0.48\textwidth]{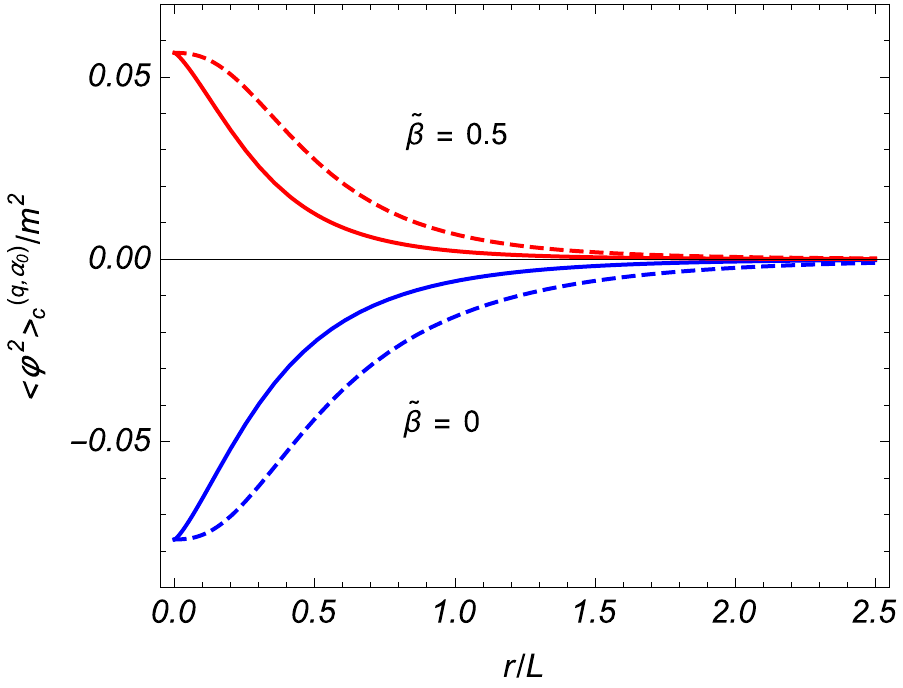}}
	\caption{VEV of the field squared for $D=3$. The left plot is the cosmic string contribution in a background without compactification in terms of the radial distance, considering $\tilde{\beta}=0.25$ and different values of the $\alpha_0$. The right plot is the contribution to the
field squared induced by the compactification along
the $z$-axis as function of $r/L$ considering $\tilde{\beta}=0$ (blue curves), $\tilde{\beta}=0.5$ (red curves), $\alpha_0=0.25$ (full curves),
$\alpha_0=0.5$ (dashed curves) and $mL=0.75$. In both plots we have $q=2.5$.}
	\label{fig01}
\end{figure}

In Fig.\ref{fig01} we have plotted the VEV of the field squared for $q=2.5$. The left plot is the contribution for
the geometry of a cosmic string with no compactification, Eq. \eqref{phi2_cs}, 
for different values of $\alpha_0$. We note that this contribution for the VEV is divergent at the origin and goes to zero with the radial distance.
The right plot of the Fig. \ref{fig01} shows the VEV of the field squared induced by the compactification and the planar angle defict, 
Eq. \eqref{phi2_decomp2}, considering $mL=0.75$. For this contribution  we have
$\alpha_0=0.25$ (full curves) and $\alpha_0=0.5$ (dashed curves). In addition,
the blue and the red curves are for $\tilde{\beta}=0$ and $\tilde{\beta}=0.5$, respectively.
This contribution, as mentioned before, has a finite value at the origin and is even function of the parameter $\tilde{\beta}$.

\section{Vacuum expectation value (VEV) of the energy-momentum tensor}
\label{Sec.EMT}
Another important characteristic of the vacuum state is the VEV of the energy–momentum tensor. To develop this quantity, we use the formula below \cite{Wagner}:
\begin{eqnarray}
\label{EMT}
\langle T_{\alpha\beta}\rangle=\lim_{x'\to x}\left(D_{\alpha'}D_\beta^*+ D_{\beta'} D_\alpha^*\right) W(x',x)-2\left[\xi R_{\alpha\beta}+\xi\nabla_\alpha\nabla_\beta-(\xi-1/4)g_{\alpha\beta}\Box\right]
\langle|\varphi|^2\rangle \  , 	
	\end{eqnarray}
where for the spacetime under consideration the Ricci tensor, $R_{\alpha\beta}$, vanishes for points outside the string. 

Similar to the case of the field squared, the VEV of the energy–momentum tensor is presented in the decomposed form:
\begin{eqnarray}
\langle T_{\alpha\beta}\rangle=\langle T_{\alpha\beta}\rangle_{cs}+\langle T_{\alpha\beta}\rangle_c \  .
\end{eqnarray}

Because the VEV of the field squared, given by \eqref{phi2_cs} and \eqref{phi2_c}, depend only on the radial coordinate, the d’Alembertian of both quantities that appears in \eqref{EMT}  are explicitly written below:
\begin{eqnarray}
\Box\langle|\varphi|^2\rangle_{cs}&=-&\frac{16m^{D+1}}{(2\pi)^{\frac{D+1}{2}}}\left\{\sum_{k=1}^{[q/2]}\cos(2k\pi \alpha_0)s_k^2\left(2m^2r^2s_k^2f_{\frac{D+3}{2}}(u_{k,0})- f_{\frac{D+1}{2}}(u_{k,0})\right)\right.\nonumber\\
&-&\frac{q}{\pi}\int_0^\infty dy \frac{g(q,\alpha_0,2y)c_y^2}{\cosh(2qy)-\cos(q\pi)}\left(2m^2r^2c_y^2f_{\frac{D+3}{2}}(u_{y,0})- f_{\frac{D+1}{2}}(u_{y,0})\right) \  ,
\end{eqnarray}
and
\begin{eqnarray}
\Box\langle|\varphi|^2\rangle_{c}&=-&\frac{32m^{D+1}}{(2\pi)^{\frac{D+1}{2}}}
\sum_{l=1}^\infty \cos(2\pi l\tilde{\beta})\left\{\sum_{k=1}^{[q/2]}\cos(2k\pi \alpha_0)s_k^2\left(2m^2r^2s_k^2f_{\frac{D+3}{2}}(u_{k,l})- f_{\frac{D+1}{2}}(u_{k,l})\right)\right.\nonumber\\
&-&\frac{q}{\pi}\int_0^\infty dy \frac{g(q,\alpha_0,2y)c_y^2}{\cosh(2qy)-\cos(q\pi)}\left(2m^2r^2c_y^2f_{\frac{D+3}{2}}(u_{y,l})- f_{\frac{D+1}{2}}(u_{y,l})\right) \  ,
\end{eqnarray}
where, for both expressions above, we adopted the notation,
\begin{eqnarray}
u_{k,l}=m\sqrt{4r^2s_k^2+(lL)^2} \ {\rm and} \  u_{y,l}=m\sqrt{4r^2c_y^2+(lL)^2} \  . 
\end{eqnarray}
For the geometry under consideration, only the differential operators $\nabla_r\nabla_r$ and $\nabla_\varphi\nabla_\varphi$ present contributions when acting on the VEV of the field squared.

The other contributions to the VEV of the energy-momentum tensor in \eqref{EMT}, comes from the differential operator acting on the Wightman function. As to the azimuthal contribution, it is preferable to apply $D_{\varphi'}^*D_\varphi$ in \eqref{eq16} following by taking the coincidence limit of the angular variable. Doing this we arrive at the summation below,
\begin{equation}
{\bar{I}}(\alpha,q,z)=\sum_{n=-\infty}^\infty q^2(n+\alpha)^2 I_{q|n+\alpha|}(z) \  , 
\label{nsum1}
\end{equation}
with $z=urr'$. This sum can be developed by using the differential equation obeyed by the modified Bessel function. So we have:
\begin{eqnarray}
{\bar{I}}(\alpha,q,z)&=&\left(z^2\frac{d^2}{dz^2}+z\frac{d}{dz}-z^2\right)
\sum_{n=-\infty}^{\infty}I_{q|n+\alpha|}(z) \ ,
\end{eqnarray}
being \cite{Braganca}
\begin{eqnarray}
\sum_{n=-\infty}^{\infty}I_{q|n+\alpha|}(z)&=&\frac{2}{q}\sideset{}{'}\sum_{k=0}^{[q/2]}\cos(2k\pi\alpha_0)e^{z\cos(2k\pi/q)}
-\frac{2}{\pi}\int_{0}^{\infty}dy\frac{e^{-z\cosh (2y)}g(q,\alpha_0,2y)}{\cosh(2qy)-\cos(\pi q)} \ .
\end{eqnarray}
The prime on the sign of the summation means that in the case $q=2p$ the term $k=q/2$ should be taken with the coefficient $1/2$, and also the term $k=0$.

The noncompactified part of the VEV of the energy–momentum tensor, $\langle T_{\alpha}^{\beta}\rangle_{cs}$,  is found from \eqref{EMT}, by making use of the expressions of the  corresponding renormalized Wightman function, as explanied in \eqref{phiren},  and the renormalized VEV of the field squared. After long but straightforward calculations, we found:
\begin{eqnarray}
\langle T_{\alpha}^{\beta}\rangle _{\text{cs}}&=&\frac{4m^{D+1}}{(2\pi )^{(D+1)/2}}\left[
\sum_{k=1}^{[q/2]}\cos(2\pi k\alpha_0) F_{\alpha,0}^{\beta}(2mr,s_{k})-\frac{q}{\pi }
\int_{0}^{\infty }dy\frac{g(q,\alpha_0,2y)}{\cosh (2qy)-\cos (q\pi
	)}\right.\nonumber\\
&\times&\left.F_{\alpha,0}^{\beta}(2mr,c_y)\right] ,  \label{EMTcs}
\end{eqnarray} 

As to the compactified part, we obtain,
\begin{eqnarray}
\langle T_{\alpha}^{\beta}\rangle _{c}&=&\frac{8m^{D+1}}{(2\pi )^{(D+1)/2}}
\sum_{l=1}^{\infty }\cos (2\pi l{\tilde\beta} )\left[ \sideset{}{'}{\sum}%
_{k=0}^{[q/2]}\cos(2\pi k\alpha_0)F_{\alpha,l}^{\beta}(2mr,s_{k})\right.\nonumber\\
&-&\left.\frac{q}{\pi }
\int_{0}^{\infty }dy\frac{g(q,\alpha_0,2y)}{\cosh (2qy)-\cos (q\pi
	)}F_{\alpha,l}^{\beta}(2mr,c_y)\right]  \ ,  \label{EMTc}
\end{eqnarray}
 where, for both expressions above, we use the notation below:
 \begin{eqnarray}
 F_{0,l}^{0}(u,v)&=&(1-4\xi)v^2[u^2v^2f_{(D+3)/2}(w_l)-2f_{(D+1)/2}(w_l)]- f_{(D+1)/2}(w_l)\nonumber\\
 F_{1,l}^{1}(u,v) &=&\left( 4\xi v^{2}-1\right) f_{(D+1)/2}(w_l),  \nonumber\\
 F_{2,l}^{2}(u,v) &=&\left( 1-4\xi v^{2}\right) \left[
 u^{2}v^{2}f_{(D+3)/2}(w_l)-f_{(D+1)/2}(w_l)\right] ,  \nonumber \\
 F_{3,l}^{3}(u,v) &=&F_{0,l}^{0}(u,v)+\left( mlL\right) ^{2}f_{(D+3)/2}(w_l) \ ,
 \label{Fij}
 \end{eqnarray}
 being
 \begin{equation}
 w_l=\sqrt{u^{2}v^{2}+\left( lmL\right) ^{2}}  \ .  \label{yl}
 \end{equation}
 The function $f_\nu(z)$ has been defined in \eqref{eq17}. For the components on the extra dimensions, one has $ F_{\alpha}^{\alpha}(u,v)= F_{0}^{0}(u,v)$ (no summation over $\alpha$).
 
\begin{figure}[h]
	\centering
	{\includegraphics[width=0.48\textwidth]{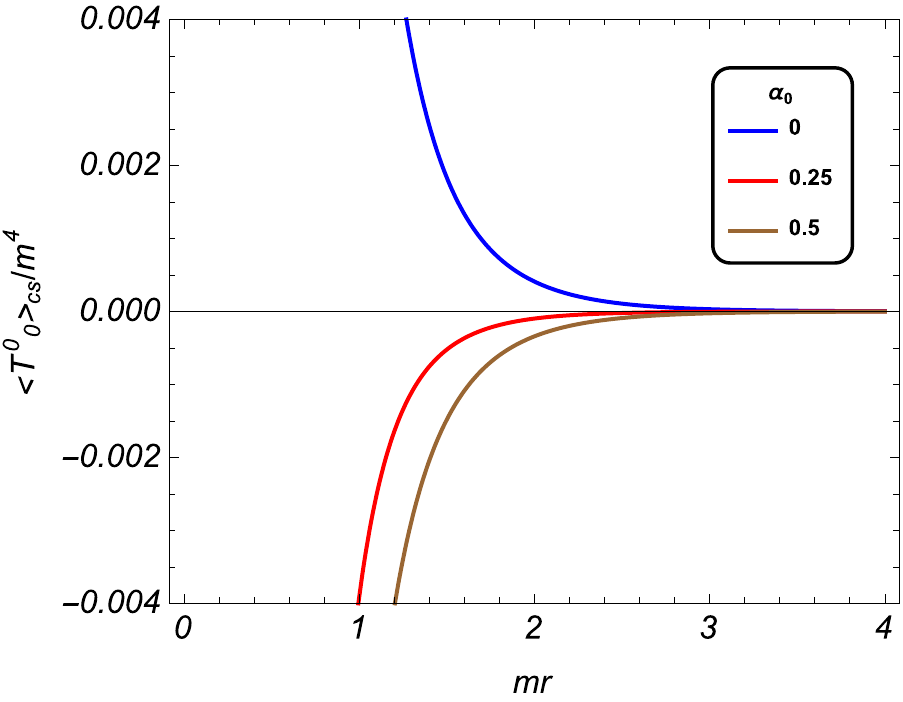}}
	\hfill
	{\includegraphics[width=0.48\textwidth]{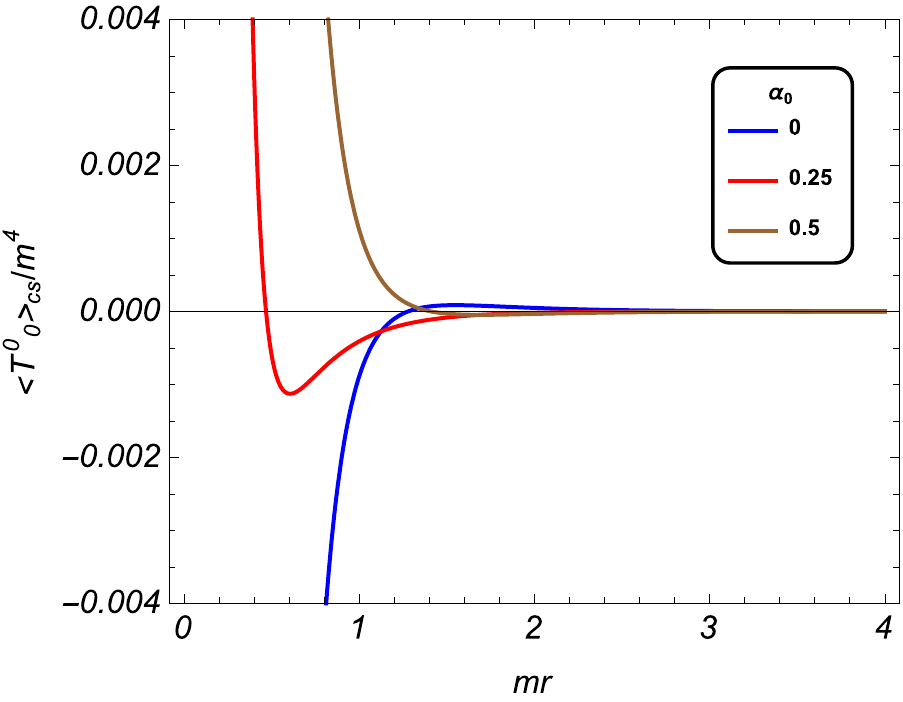}}
	\caption{The VEV of the energy density in the uncompactified cosmic string geometry, $\langle T_0^0\rangle_{cs}/m^4$, for different values of $\alpha_0$, $q=2.5$ and for $D=3$ as function of $mr$. The left plot is for $\xi=0$ (minimal coupling) while the right one is for $\xi=1/6$ (conformal coupling).}
	\label{fig02}
\end{figure}

 In  Fig. \ref{fig02} we plot the VEV of the energy density in units of ``$m^4$'' for different values of the parameter $\alpha_0$ for an uncompactified cosmic string contribution for the minimal (left plot) and the conformal (right plot) coupling, considering $q=2.5$. This contribution is divergent at the string's axis $(r=0)$ and goes to zero at the large distance from the string.

For the pure cosmic string contribution and considering a massless scalar field, from the equation \eqref{EMTcs} we found
\begin{eqnarray}
\langle T_{\alpha}^{\beta}\rangle _{\text{cs}}&=&\frac{2\Gamma((D+1)/2)}{(4\pi )^{(D+1)/2}r^{D+1}}\left\{
\left[\frac{D-1}{D}g_D(q,\alpha_0)-g_{D+2}(q,\alpha_0)\right] \rm{diag}(1,1,-D,1,...,1)\right.\nonumber\\
&-&\left.4(D-1)(\xi-\xi_D)g_D(q,\alpha_0)\left(1,\frac{-1}{D-1},\frac{D}{D-1},1,...,1\right)\right\},
\label{EMTcsmassless}
\end{eqnarray}
where the function $g_D(q,\alpha_0)$ have been defined by \eqref{gD}.

Considering the case of a massless scalar field, the equation \eqref{EMTc} becomes
\begin{eqnarray}
\langle T_{\alpha}^{\beta}\rangle _{c}&=&\frac{4\Gamma((D+1)/2)}{\pi^{(D+1)/2}L^{D+1}}
\sum_{l=1}^{\infty }\frac{\cos (2\pi l{\tilde\beta} )}{l^{D+1}}\left[ \sideset{}{'}{\sum}%
_{k=0}^{[q/2]}\frac{\cos(2\pi k\alpha_0)}{[(2rs_k/lL)^2+1]^{(D+3)/2}}F_{\alpha,l}^{(0)\beta}(2r/lL,s_{k})\right.\nonumber\\
&-&\left.\frac{q}{\pi }
\int_{0}^{\infty }dy\frac{g(q,\alpha_0,2y)}{\cosh (2qy)-\cos (q\pi
	)}\frac{F_{\alpha,l}^{(0)\beta}(2r/lL,c_y)}{[(2rc_y /lL)^2+1]^{(D+3)/2}}\right]  \ ,  \label{EMTcmassless}
\end{eqnarray}
with the notations below:
\begin{eqnarray}
F_{0,l}^{(0)0}(u,v)&=&(1-4\xi)v^2[(D-1)u^2v^2-2]-u^2v^2-1\nonumber\\
F_{1,l}^{(0)1}(u,v) &=&\left( 4\xi v^{2}-1\right) (u^2v^2+1),  \nonumber\\
F_{2,l}^{(0)2}(u,v) &=&\left( 1-4\xi v^{2}\right)(Du^2v^2-1) ,  \nonumber \\
F_{3,l}^{(0)3}(u,v) &=&F_{0,l}^{(0)0}(u,v)+D+1 \ ,
\label{Fijmassless}
\end{eqnarray}

Similarly to the calculation of the field squared, the component $\langle T_0 ^0 \rangle_c$ of the EMT tensor
can also be written in a decomposed form as
\begin{equation}
\langle T_0 ^0 \rangle_c=\langle T_0 ^0 \rangle_c ^{(0)}+
\langle T_0 ^0 \rangle_c^{(q,\alpha_0)}.
\end{equation}

As before, we have a contribution which is a pure tolopological term, given by the first term on the right hand side of the above equation. It does not dependent on $q$ and $\alpha_0$, and corresponds to the $k=0$ term of the summation. The other contribution dependes on the magnetic flux and planar angle deficit.

The expression for first term is presented as
\begin{equation}
\label{EMTtopo}
\langle T_0 ^0 \rangle_c^{(0)}=-\frac{4m^{D+1}}{(2\pi)^{\frac{D+1}{2}}}\sum_{l=1}^{\infty}\cos(2\pi l\tilde{\beta})f_{\frac{D+1}{2}}(lmL).
\end{equation}
In the limit of $mL\gg 1$, the main contribution of the above equation comes from the $l=1$ term and is written as
\begin{equation}
\langle T_0 ^0 \rangle_c^{(0)}=-\frac{2}{(2\pi m)^{D/2}}\frac{\cos(2\pi\tilde{\beta})}{L^{\frac{D+2}{2}}}e^{-mL}.
\end{equation}

For the second term, $\langle T_0 ^0 \rangle_c^{(q,\alpha_0)}$, we have:
\begin{eqnarray}
\langle T_0 ^0 \rangle_c^{(q,\alpha_0)}&=&\frac{8m^{D+1}}{(2\pi )^{(D+1)/2}}
\sum_{l=1}^{\infty }\cos (2\pi l{\tilde\beta} )\left[ \sum_{k=1}^{[q/2]}\cos(2\pi k\alpha_0)F_{0,l}^{0}(2mr,s_{k})\right.\nonumber\\
&-&\left.\frac{q}{\pi }
\int_{0}^{\infty }dy\frac{g(q,\alpha_0,2y)}{\cosh (2qy)-\cos (q\pi
	)}F_{0,l}^{0}(2mr,c_y)\right]  \ .
	\label{EMTctopo2}
\end{eqnarray}

Now we want to investigate the behavior of the energy-density induced by the compactification at the string's core, $r=0$. In the case where there is no magnetic flux running along the string, this term is finite as we have shown in \cite{Mello12}; however the presence of the magnetic flux changes this situation. The main reason is in the expression for the function $g(q,\alpha_0,2y)$, Eq. \eqref{gfunction}, that appears in the integral. Being $\alpha_0=0$, there are no hyperbolic cosines involving the variable of integration, $y$, and the integral is finite. In addition to this function, the integrand also presents a term $F_{0,l}^{0}(u,v)$, given by the first expression in \eqref{Fijmassless}. In fact the integral contribution in \eqref{EMTctopo2} is responsible for the most relevant  information. In order to analyze the finiteness of this integral, we have to analyze its integrand for large value of $y$.  In this limit we can  approximate $\cosh(y)\approx e^y/2$. So, by adopting this procedure, we can conclude that the integral is finite only if $q>1/|\alpha_0|$; so, the value of $\langle T_0 ^0 \rangle_c^{(q,\alpha_0)}$ at origin can be obtained by taking directly $r=0$. For $q<1/|\alpha_0|$ the integral is divergent.  Taking all these informations in consideration, and after some intermediate steps and using the integral representation for the Macdonald function, Eq. \eqref{Kfunction}, we could find that the leading divergence term of the Eq. \eqref{EMTctopo2} is given by
\begin{eqnarray}
\langle T_0^0\rangle_c^{(q,\alpha_0)}&=&-\frac{8q(1-4\xi)\sin(q\pi\alpha_0)m^{D+1}}{\pi^{1+D/2}(mr)^{2(1-q\alpha_0)}}\sum_{l=1}^\infty\cos(2\pi l\tilde{\beta})\nonumber\\
&&\times\left[f_{(D/2)-2(1-q\alpha_0)}(lmL\sqrt{2})
-f_{(D/2)+1-2q\alpha_0}(lmL\sqrt{2})\right].
\label{EMTdiv}
\end{eqnarray}
Although we have not shown explicitly, other components of the EMT also are expected to have a divergent behaviour at the string's core in the case of $q<1/|\alpha_0|$.

\begin{figure}[h]
	\centering
	{\includegraphics[width=0.49\textwidth]{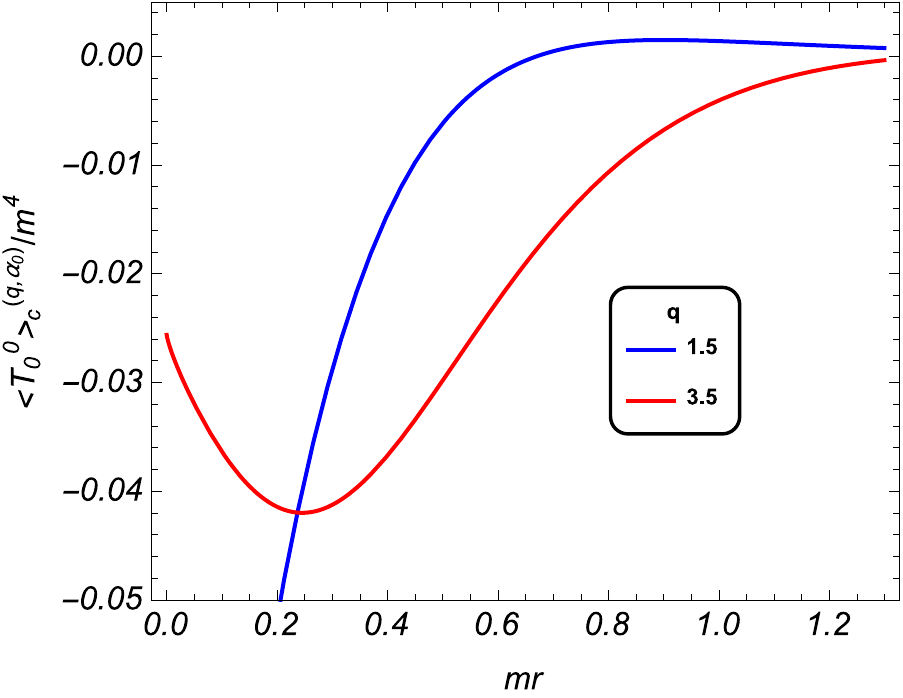}}
	\hfill
	{\includegraphics[width=0.49\textwidth]{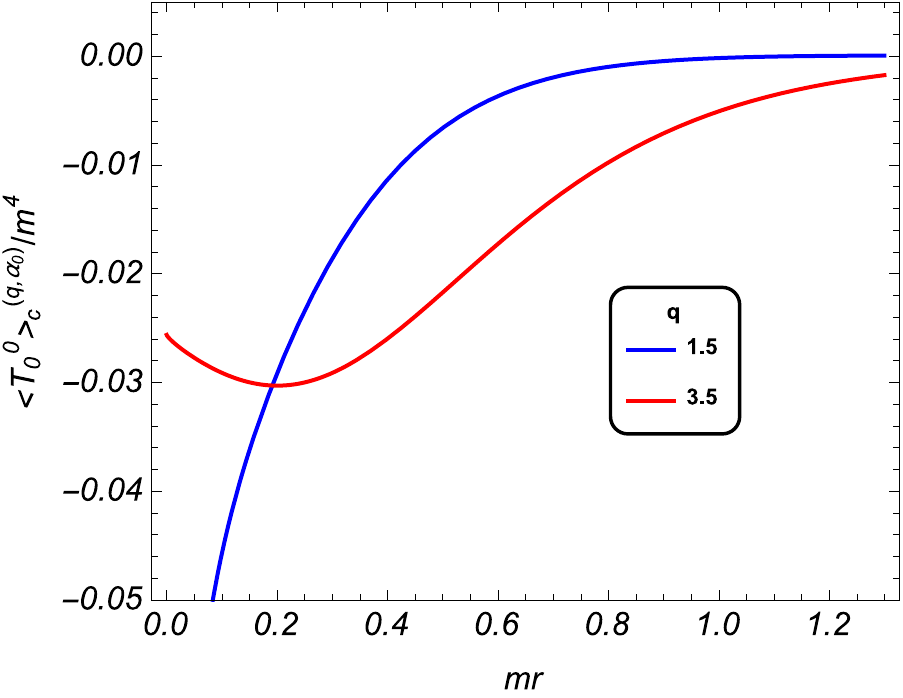}}
	\caption{The behaviors of the VEV of the energy density induced by the compactification, the magnetic flux and the planar angle deficit, $\langle T_0^0\rangle_{c}^{(q,\alpha_0)}/m^4$ as function of $mr$ are exhibited  for the minimally (left plot) and conformally (right plot) coupled scalar fields. The plots are  for different values of $q$, $\alpha_0=0.4$, $\tilde{\beta}=0.25$, $mL=0.75$ and $D=3$.}
	\label{fig03}
\end{figure}

In  Fig. \ref{fig03} we plot the VEV of the energy density induced by the compactification considering the contributions of the magnetic flux and the planar angle deficit, Eq. \eqref{EMTctopo2}, as function of $mr$ for different values of the parameter $q$ and fixing $\tilde{\beta}=0.75$ and $\alpha_0=0.4$.  In these plots are considered the minimally (left plot) and conformally (right plot) coupled scalar fields. Note that, as mentioned before, for the case where $q=3.5$ the behaviors of the energy density, for both values of $\xi$, are finite at $r=0$, while for $q=1.5$ they diverge.

Considering $mL\gg1$ the main contribution for the Eq.  \eqref{EMTctopo2} comes from the $l=1$ term which, is given by
\begin{eqnarray}
\langle T_0 ^0 \rangle_c^{(q,\alpha_0)}&\approx&
-\frac{\cos (2\pi {\tilde\beta} )e^{-mL}}{2^{\frac{D-5}{2}}L^{\frac{D+2}{2}}}\left(\frac{m}{\pi}\right)^{D/2}
\left\{\sum_{k=1}^{[q/2]}\cos(2\pi k\alpha_0)\left[2(1-4\xi)s^2_k+\frac{1}{\sqrt{2}}\right]\right.\nonumber\\
&-&\left.\frac{q}{\pi }
\int_{0}^{\infty }dy\frac{g(q,\alpha_0,2y)}{\cosh (2qy)-\cos (q\pi
	)}\left[2(1-4\xi)c_y^2+\frac{1}{\sqrt{2}}\right]\right\}  \ .
\label{E00C}
\end{eqnarray}

\begin{figure}[h]
	\centering
	{\includegraphics[width=0.47\textwidth]{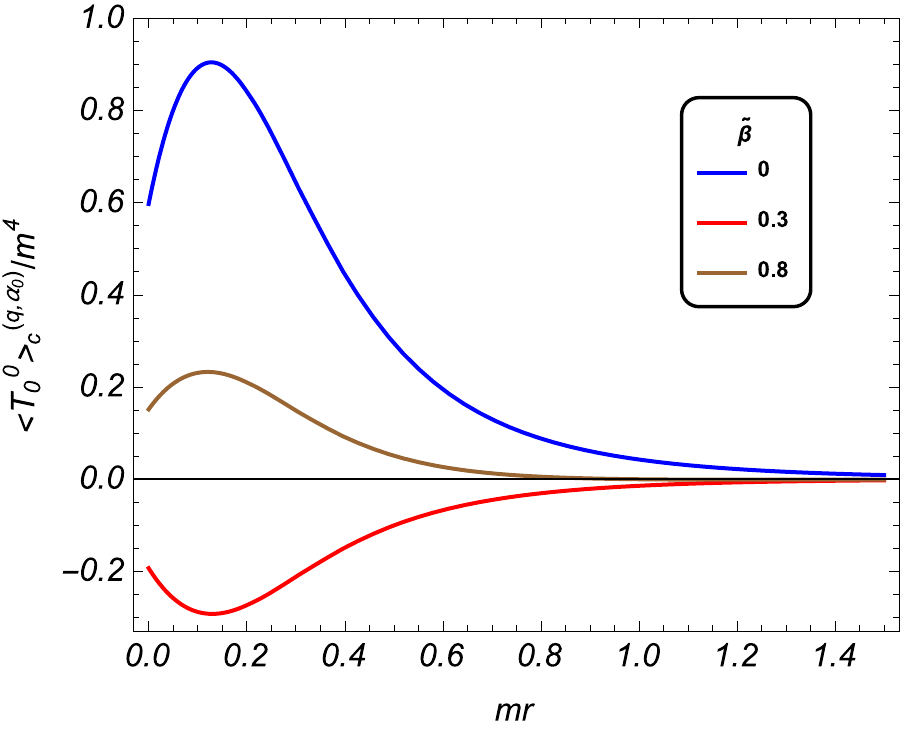}}
	\hfill
	{\includegraphics[width=0.48\textwidth]{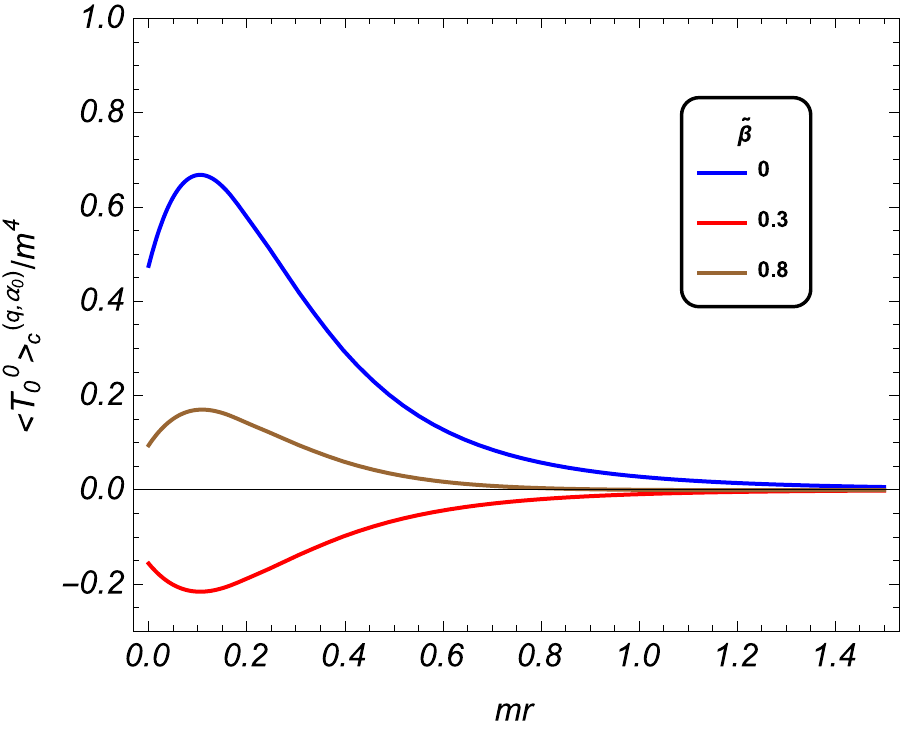}}
	\caption{The VEV of the energy density induced by the compactification, the magnetic flux and the planar angle deficit, $\langle T_0^0\rangle_{c}^{(q,\alpha_0)}/m^4$ as function of $mr$. We consider different values of $\tilde{\beta}$, $\alpha_0=0.4$, $q=3.5$, $mL=0.75$ and $D=3$. The left plot is for $\xi=0$ while the right one is for $\xi=1/6$.}
	\label{fig04}
\end{figure}

 In  Fig. \ref{fig04} we have the plot of the energy density induced by the compactification as function of $mr$ considering different values of the parameter $\tilde{\beta}$ for $\alpha_0=0.4$ and $q=3.5$, and also having the minimally and conformally coupled scalar cases.\footnote{The values adopted for the parameters $q$ and $\alpha_0$ ensures finite values of the energy density at the string's core} By these plots we can infer that the intensity and behavior of the energy density depend crucially on the parameter 
 $\tilde{\beta}$.  

Finally, it can be proved that both contributions to the VEV of the energy-momentum tensor satisfy the covariant conservation condition, $\nabla_\beta \langle T_\alpha ^\beta \rangle=0$, which for the geometry under consideration reduces itself to $\langle T^\phi_\phi\rangle=\partial_r\langle T^r_r\rangle$. In addition, they also obey the trace relation
\begin{equation}
\label{EMTtrace}
\langle T	_\alpha ^\alpha \rangle= 2\left[D(\xi-\xi_D)\nabla_\alpha \nabla^\alpha \langle| \varphi|^2\rangle
+m^2\langle| \varphi|^2\rangle\right] \  .
\end{equation}
In particular, it is traceless for a conformally coupled massless field.

\section{Conclusions}
\label{Concl}
In this paper, we have investigated the VEV of the energy-momentum tensor associated with a charged scalar quantum field
considering a high-dimensional compactified cosmic string spacetime. We considered the presence of a magnetic flux running through the string's
 core and another one running along the center of the coordinate $z$ which is compactified to a circle. As the first step, we evaluate the
positive frequency of the Wightman function  and found a closed form, for a general value of the parameter $q$, Eq. \eqref{wightmanfinal}.
We found that this function can be decomposed in two contributions: one due to a geometry of a cosmic string without compactification
and another one induced by the compactification itself. With this function in hands, we were able to evaluate the renormalized VEV of the field squared taking the limit of coincidence of the arguments in the Eq. \eqref{wightmanfinal} eliminating the Minkowiskian part. The VEV of the field squared also could be decomposed in a contribution due
a geometry of cosmic string without compactification, Eq. \eqref{phi2_cs}, and another one due the compactification
along the $z$-axis, Eq. \eqref{phi2_c}.  Expressions for the massless field and in the limit of $mr\gg1$ were calculated, Eqs. \eqref{phi2massless}
and \eqref{phi2mr}, respectively, for $\langle |\varphi|^2\rangle_{cs}$. The part of the field
squared induced by the compactification, $\langle |\varphi|^2\rangle_{c}$, also
could be decomposed in two contributions, Eq. \eqref{phi2_decomp}. The first one, Eq. \eqref{phi0}, being a pure topological term dependent only on the
compactification along the $z$-axis, which is the $k=0$ term of the summation over $k$ in the Eq. \eqref{phi2_c}.
The second contribution, Eq. \eqref{phi2_decomp2}, presents a dependence on the magnetic flux and the
planar angle deficit. For both expressions we investigated the limit in which $mL>>1$, Eqs. \eqref{Phi_1} and \eqref{Phi_2}, respectively. As expected both expressions go to zero in this limit. Also we have presented in Fig. \ref{fig01} the behaviour of the VEV of the field squared as function of $mr$ for the uncompactified term, and $r/L$ for the compactified one.

Another analysis developed in this paper was the VEV of the energy-momentum tensor making use of the formula Eq. \eqref{EMT}. Similarly to the case
of the Wightman function and of the field squared, we decomposed the components of the energy-momentum tensor in
contributions due the geometry of a cosmic string with no compactification and in a contribution
induced by the compactification, the magnetic flux and the planar angle deficit, Eqs. \eqref{EMTcs} and \eqref{EMTc}.
Considering the VEV of the energy density, we have presented expressions for the case of a massless scalar field, Eqs. \eqref{EMTcsmassless}
and \eqref{EMTcmassless}. The cosmic string part of the energy density presents a divergence at the string's axis, $r=0$, and goes to zero for large distances from the string, as it can be seen in Fig. \ref{fig02}. The part induced by the compactification could be decomposed in two contributions: one dependending only on the compactification, which is a pure topological
term, Eq. \eqref{EMTtopo}, and another one which is induced by the planar angle deficit and the magnetic flux,
Eq. \eqref{EMTctopo2}. We have evaluated both expressions in the limit of $mL>>1$ and we found for both exponential decay behaviors. Another point that was investigated concerns the behavior of the energy-density induced by the planar angle deficit and the magnetic flux, Eq. \eqref{EMTctopo2}.  We noticed that, due to the presence of the magnetic flux running along the string, this part presents a divergence at the string's core if $q<1/|\alpha_0|$. For the case where $q>1/|\alpha_0|$ this contribution has a finite value at the origin. These facts were exemplified in the plots of Fig. \ref{fig03}. Moreover, we explicitly calculated the leading divergence term which is given by the Eq.
\eqref{EMTdiv}. We also analyzed  graphically,  the dependence of \eqref{EMTctopo2} with the parameter 
$\tilde{\beta}$. By the Fig. \ref{fig04} we can infer that both, intensity and the sign, depend on this parameter.  

Finally we have checked that the  VEV of the energy-momentum tensor that we have calculated satisfies the covariant conservation condition
and obeys a trace relation given by the Eq. \eqref{EMTtrace}. 
\section*{Acknowledgments}
E.A.F.B would like to thank the Brazilian agencies CAPES and CNPq for financial support. H.F.S.M and E.R.B.M are partially supported by the Brazilian agency CNPq under grants 305379/2017-8 and 313137/2014-5, respectively.


\providecommand{\href}[2]{#2}\begingroup\raggedright

\endgroup
\end{document}